\documentclass[prl,twocolumn,superscriptaddress]{revtex4}

\usepackage{graphicx}
\usepackage{dcolumn}
\usepackage{bm}

\begin{document}

\preprint{ATB-1}

\title{Spin Dynamics in Stripe-Ordered
La$_{5/3}$Sr$_{1/3}$NiO$_4$}

\author{A.T. Boothroyd}
\email{a.boothroyd1@physics.ox.ac.uk}
\homepage{http://xray.physics.ox.ac.uk/Boothroyd}\affiliation{
Department of Physics, Oxford University, Oxford, OX1 3PU, United
Kingdom }
\author{D. Prabhakaran}\affiliation{
Department of Physics, Oxford University, Oxford, OX1 3PU, United
Kingdom }
\author{P.G. Freeman}\affiliation{
Department of Physics, Oxford University, Oxford, OX1 3PU, United
Kingdom }
\author{S.J.S. Lister}\altaffiliation[Present address: ]{ Oxford Magnet
Technology Ltd, Witney, Oxfordshire, OX29 4BP, United
Kingdom}\affiliation{ Department of Physics, Oxford University,
Oxford, OX1 3PU, United Kingdom }
\author{M. Enderle}\affiliation{
Institut Laue-Langevin, BP 156, 38042 Grenoble Cedex 9, France }
\author{A. Hiess}\affiliation{
Institut Laue-Langevin, BP 156, 38042 Grenoble Cedex 9, France }
\author{J. Kulda}\affiliation{
Institut Laue-Langevin, BP 156, 38042 Grenoble Cedex 9, France }

\date{\today}

\begin{abstract}
Polarized and unpolarized neutron inelastic scattering has been
used to measure the spin excitations in the spin-charge-ordered
stripe phase of La$_{5/3}$Sr$_{1/3}$NiO$_4$. At high energies,
sharp magnetic modes are observed characteristic of a static
stripe lattice. The energy spectrum is described well by a linear
spin wave model with intra- and inter-stripe exchange interactions
between neighbouring Ni spins given by $J=15\pm1.5$\,meV and
$J'=7.5\pm1.5$\,meV respectively. A pronounced broadening of the
magnetic fluctuations in a band between $10$\,meV and $25$\,meV is
suggestive of coupling to collective motions of the stripe domain
walls.
\end{abstract}

\maketitle

Over the past decade, the tendency of doped antiferromagnetic
oxides to exhibit symmetry-broken phases involving the ordering of
both spin and charge has become increasingly apparent. Widespread
interest was generated by the discovery of a stripe-like,
spin--charge ordered phase in a non-superconducting layered
cuprate \cite{Tranquada-Nature-1995}. This stripe phase, the like
of which has now been found in many other doped transition metal
oxide systems, is characterized by parallel lines of holes that
act as charged domain walls separating regions of
antiferromagnetically-ordered spins. Its discovery fuelled a
debate about the role played by stripe correlations in the
formation of the superconducting state in the cuprates
\cite{SC-stripes-theory}, and stimulated numerous experimental
investigations into stripe pehenomena. These have focussed
principally on the La$_{2-x}$Sr$_x$NiO$_{4+\delta}$ series, which
exhibits stripe ordering over a wide range of hole concentration
\cite{stripes-nickelates-expt, Yoshizawa-PRB-2000}.

While the static properties of ordered stripe phases are now quite
well characterized, less is known about their dynamics.
Specifically, there is a lack of information on how the collective
motions of the holes in a stripe domain wall couple to the spin
dynamics of the antiferromagnetic domains. One way to make
progress in understanding stripe dynamics is to use neutron
inelastic scattering to probe the spin excitation spectrum as a
function of wavevector and energy in simple compounds exhibiting
well-correlated stripe ordering. As well as providing information
on the microscopic interactions that govern the properties of
stripes, such studies can address the question of whether stripes
are essential or incidental to the mechanism of superconductivity.

\begin{figure}
\includegraphics{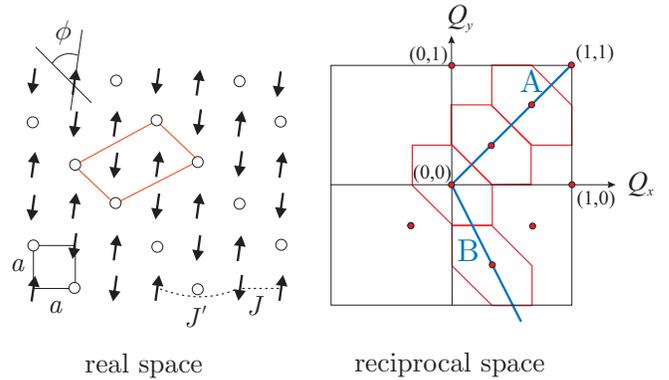}
\caption{(a) Model for the stripe order found in the Ni--O plane
of La$_{5/3}$Sr$_{1/3}$NiO$_4$ \protect\cite{Yoshizawa-PRB-2000,
Lee-PRB-2001}. Arrows denote $S=1$ spins on the Ni$^{2+}$ sites,
open circles represent Ni$^{3+}$ holes. The O sites are not shown.
$\phi$ is the angle between the spin axes and the stripe
direction. Primitive unit cells of the NiO$_2$ square lattice and
stripe superstructure are indicated respectively by the small
square and parallelogram. $J$ and $J'$ are respectively the intra-
and inter-stripe exchange interactions between nearest-neighbour
Ni spins. (b) Diagram of reciprocal space showing several
Brillouin zones for the stripe-order superlattice shown in (a).
The $(h,k)$ indices correspond to the NiO$_2$ square lattice, and
the small circles are the stripe superlattice zone centres. The
lines marked A and B indicate the directions of the scans used in
the experiment. \label{fig:1} }
\end{figure}

Here we report polarized- and unpolarized-neutron scattering
measurements of the spin excitation spectrum in the ordered stripe
phase of La$_{5/3}$Sr$_{1/3}$NiO$_4$. This composition has not
been studied before by neutron inelastic scattering, and was
chosen because the spin--charge order is particularly well
correlated and has a very simple superstructure commensurate with
the crystal lattice \cite{Yoshizawa-PRB-2000, Cheong-PRB-1994,
Du-PRL-2000, Lee-PRB-2001} --- see Fig.\,\ref{fig:1}(a). Compared
with earlier neutron scattering studies of stripe-ordered
La$_{2-x}$Sr$_x$NiO$_{4+\delta}$ compounds
\cite{INS-stripes-low-energy, Bourges-cond-mat-2002} our
measurements cover a wider range of wavevector and energy than
previously explored in any one compound, and the simplicity of the
magnetic structure of La$_{5/3}$Sr$_{1/3}$NiO$_4$ allows us to
perform a quantitative analysis of the data by linear spin-wave
theory. Thus, for the first time in a stripe-ordered compound, we
determine values for the nearest-neighbour coupling strengths for
spins within a stripe domain and for spins separated by a charged
domain wall. We also observe an anomaly in the magnetic scattering
between $10$\,meV and $25$\,meV that could, we suggest, originate
from coupling between spin excitations and collective motions of
the charged domain walls.

Two single crystals of La$_{5/3}$Sr$_{1/3}$NiO$_4$ in the form of
rods, 7--8\,mm in diameter and $\sim$40\,mm in length, were used
for the experiments. Both were grown from the same batch of
starting powder by the floating-zone method
\cite{Prabhakaran-JCG-2002}. Their oxygen excess was determined by
thermogravimetric analysis to be $\delta=0.01\pm0.01$, and their
electrical and magnetic properties were found to be in very good
agreement with data in the literature.

Neutron inelastic scattering measurements were performed on the
IN20 and IN22 triple-axis spectrometers at the Institut
Laue-Langevin. The incident and final neutron energy was selected
by Bragg reflection from an array of either pyrolytic graphite
(PG) or Heusler alloy crystals, depending on whether unpolarized
or polarized neutrons were employed. Scans were performed with a
fixed final energy of either $E_f = 14.7$\,meV or $34.8$\,meV, and
a PG filter was present to suppress higher-order harmonics in the
scattered beam. Two settings of the crystal were used, giving
access to the $(h, h, l)$ and $(h, -2h, l)$ planes in reciprocal
space. In this paper the reciprocal lattice is indexed with
respect to a body-centred tetragonal lattice with cell parameters
$a=3.8\,{\rm \AA}$ and $c=12.7\,{\rm \AA}$.

Long-range ordering of holes and spins in
La$_{5/3}$Sr$_{1/3}$NiO$_4$ occurs below $T_{\rm CO}\simeq 240$\,K
and $T_{\rm SO}\simeq 200$\,K respectively. The stripe arrangement
formed in the Ni--O layers is shown in Fig.\,\ref{fig:1}(a)
\cite{Yoshizawa-PRB-2000, Lee-PRB-2001}. In this model the spins
are collinear, and $\phi$ is the angle between the spin axes and
the stripe direction. All the measurements reported here were
taken at $T = 14$\,K, at which temperature $\phi = 53^\circ$
\cite{Lee-PRB-2001} and the in-plane and out-of-plane spin and
charge order correlation lengths are several hundred $\rm \AA$ and
20--50\,$\rm \AA$ respectively \cite{Yoshizawa-PRB-2000,
Du-PRL-2000}. In inelastic measurements the $c$ axis correlations
were found to decay very rapidly with energy, entirely
disappearing above 5\,meV. Over the energy range considered here
it is therefore reasonable to treat the spin dynamics as
two-dimensional.

Fig.\,\ref{fig:1}(b) shows several two-dimensional Brillouin zones
of the stripe order. We probed the excitation spectrum either by
scanning the neutron scattering vector $\bf Q$ along the lines
marked A and B at a fixed energy, or by scanning the energy at a
fixed $\bf Q$. In reality, the ordered stripe phase contains equal
proportions of two twins with stripes at $90^\circ$ to one
another. The reciprocal space for the second twin is rotated by
$90^\circ$ with respect to that shown in Fig.\,\ref{fig:1}(b).
Experimentally, one observes a superposition of the scattering
from both twins.

\begin{figure}
\includegraphics{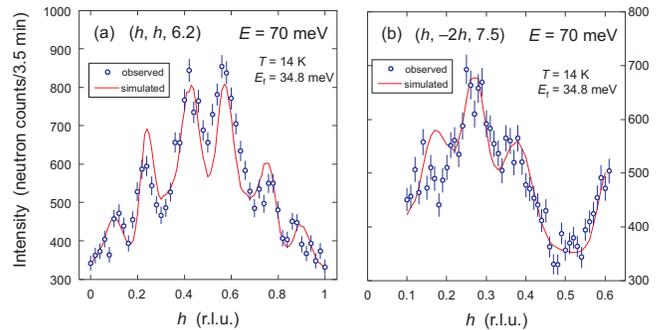}
\caption{$\bf Q$ scans parallel to (a) $(h, h, 0)$ and (b) $(h,
-2h, 0)$ at a constant energy of 70\,meV, showing the magnetic
scattering from La$_{5/3}$Sr$_{1/3}$NiO$_4$ The lines are
simulations of the scans generated by convolution of the
calculated spectrometer resolution with the spin-wave model
discussed in the text. \label{fig:2} }
\end{figure}
Assuming the ordered ground state in Fig.\,\ref{fig:1} one expects
to observe magnetic excitations dispersing from the magnetic zone
centres, and indeed this is confirmed by our measurements. To
illustrate the results, Fig.\,\ref{fig:2} shows wavevector scans
parallel to the $(h, h, 0)$ and $(h, -2h, 0)$ directions measured
with unpolarized neutrons at a fixed energy of 70\,meV. The six
peaks observed in Fig.\,\ref{fig:2}(a) correspond to spin
excitations propagating in a direction perpendicular to the
stripes away from successive zone centres along the A line in
Fig.\,\ref{fig:1}(b). Similarly, the two strongest peaks in
Fig.\,\ref{fig:2}(b) are spin excitations propagating away from
$(1/3, -2/3)$ along line B, approximately parallel to the stripes.
Under the experimental conditions used we were not able to resolve
the separate peaks associated with each zone centre when the
energy was below 37\,meV.

\begin{figure}
\includegraphics{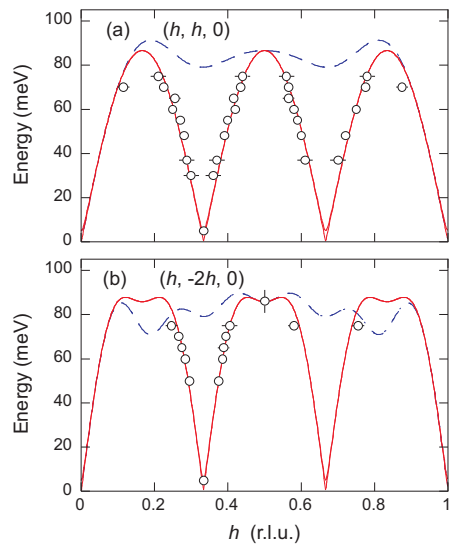}
\caption{Dispersion of the magnetic excitations in
La$_{5/3}$Sr$_{1/3}$NiO$_4$ parallel to (a) $(h, h, 0)$ and (b)
$(h, -2h, 0)$. The full and broken lines are the spin-wave model
dispersion for the two twins, calculated with parameters
$J=15$\,meV, $J'=7.5$\,meV, and $K_c = 0.07$\,meV. \label{fig:3} }
\end{figure}

In Figs.\,\ref{fig:3} we plot the dispersion of the spin
excitations along the $(h, h, 0)$ and $(h, -2h, 0)$ directions as
deduced from a series of (mainly) constant-energy scans like those
of Figs.\,\ref{fig:2}. To arrive at the data points shown in
Figs.\,\ref{fig:3} we corrected the observed peak positions for
small shifts (10--20\,\%) caused by the spectrometer resolution as
it intersects the curved dispersion surface. We determined the
corrections by convolution of the resolution function with the
model dispersion surface to be described later.

In the energy range 10--30\, meV very strong scattering from
phonons makes it difficult to identify unambiguously the magnetic
scattering with unpolarized neutrons. Therefore, to explore this
energy range we employed polarization analysis, keeping the
neutron polarization direction $\bf P$ parallel to $\bf Q$ by
adjusting currents in a Helmholtz coil mounted around the sample
position. In this configuration the neutron spins are flipped when
scattered by magnetic excitations, but remain unchanged in
non-magnetic (e.g. phonon) scattering processes.

\begin{figure}
\includegraphics{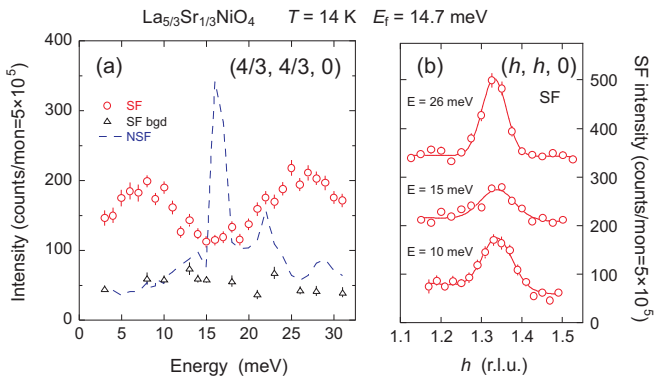}
\caption{(a) Spin-flip (SF) and non-spin-flip (NSF) scattering as
a function of energy at a constant wavevector of $(4/3, 4/3, 0)$.
The sharp peaks in the NSF channel are due to phonons. The SF
background is estimated from data collected at $(1.2, 1.2, 0)$.
(b) SF scattering intensity plotted as a function of wavevector
for energies of 10\,meV, 15\,meV and 26\,meV. The upper two scans
have been shifted vertically by 150 and 300 counts respectively.
The lines are fits to Gaussian peak profiles. A monitor of
$5\times10^5$ corresponds to a counting time of 5--8\,mins,
depending on energy. \label{fig:4} }
\end{figure}

Fig.\,\ref{fig:4}(a) shows the neutron spin-flip (SF) and
non-spin-flip (NSF) scattering observed as a function of energy
with $\bf Q$ fixed at the $(4/3, 4/3, 0)$ magnetic zone centre.
The separation of the magnetic and non-magnetic scattering by the
polarization analysis is exemplified by the sharp phonon peaks
centred on 16.5\,meV and 22\,meV in the NSF channel that are not
present in the SF channel. However, the most remarkable feature in
this scan is the broad valley centred near 15\,meV in the SF
scattering. By comparison with the SF measurements made at ${\bf
Q} = (1.2, 1.2, 0)$, where the magnetic scattering has decreased
to background level, we see that the amplitude of the magnetic
signal above background is nearly a factor 3 smaller at $E =
15$\,meV than at either $E = 10$\,meV or $E = 26$\,meV. Similar
scans measured at $(1/3, 1/3, l)$, $l = 4.5, 5.5$, show the same
valley feature. A secondary feature in Fig.\,\ref{fig:4}(a) is the
fall in magnetic intensity with energy observed below
$\sim$7\,meV.

One possible origin for the features just described could be spin
anisotropy gaps in the excitation spectrum. Below an anisotropy
gap a component of the spin fluctuations freezes out, causing a
reduction in scattering intensity. We checked this possibility by
comparing the magnetic scattering intensities for neutron
polarizations ${\bf P}\parallel{\bf Q}$ and ${\bf P}\perp{\bf Q}$.
The data at $10$\,meV, $15$\,meV and $26$\,meV were all consistent
with the assumption of isotropic transverse spin fluctuations
about the direction of the ordered moment. At $3$\,meV, however,
the signal is predominantly due to in-plane transverse spin
fluctuations. Therefore, we conclude that the drop in intensity
below $\sim$7\,meV is caused by an out-of-plane anisotropy gap,
but that the main valley feature cannot be explained by spin
anisotropy.

Fig.\,\ref{fig:4}(b) shows the SF scattering intensity measured in
a series of constant-energy scans in the neighbourhood of the
valley feature. These data show that, in addition to a reduction
in amplitude, there is also an increase in width. The $15$\,meV
peak is found to be $\sim$$50$\,\% broader in wavevector than the
$10$\,meV and $26$\,meV peaks. This broadening suggests that the
valley feature is associated with the decay of antiferromagnetic
spin excitation modes through hybridisation with another type of
excitation. One possibility is a coupling to optic phonons
intrinsic to the material, but we do not believe this to be the
case because (a) the valley region is $\sim$$10$\,meV wide, much
larger than the width of individual phonon excitations, and (b) we
do not observe any structure in the SF scattering channel due to
the excitation of phonons through a magnetoelastic interaction.
Instead, we suggest that a probable explanation for the valley
feature is the interaction between spin excitations and motions of
the stripe domain walls with characteristic frequencies in the
3--5\,THz range.

Having established the overall features of the spin excitation
spectrum we now describe the spin-wave model used to extract the
microscopic exchange parameters. The model is based on the spin
Hamiltonian
\begin{equation}
H = J\hspace{-5pt}\sum_{<ij> \atop{\rm{intra-} \atop{\rm
stripe}}}\hspace{-5pt}{\bf S}_i\cdot{\bf S}_j +
J'\hspace{-5pt}\sum_{<ij'> \atop{\rm{ inter-}\atop{\rm
stripe}}}\hspace{-5pt}{\bf S}_i\cdot{\bf S}_{j'} +
K_c\sum_i(S_i^z)^2,\label{eq_1}
\end{equation}
where the first two summations are over pairs of
nearest-neighbouring Ni spins, the first sum having both spins
within the same stripe domain and the second having the two spins
in adjacent domains separated by a line of holes. $J$ and $J'$ are
the corresponding exchange parameters --- see
Fig.\,\ref{fig:1}(a). The third term describes the out-of-plane
anisotropy. The in-plane anisotropy is much smaller than the
out-of-plane anisotropy, and so is neglected.

We calculated the energy dispersion and scattering cross-section
using the linear spin-wave approximation, and after convolving the
model spectrum with the spectrometer resolution we compared the
results with the measured scans. The parameter values giving the
best agreement with the totality of data are $J=15\pm1.5$\,meV,
$J'=7.5\pm1.5$\,meV and $K_c = 0.07\pm0.01$\,meV. For reference,
the $J$ and $K_c$ parameters obtained by Nakajima {\it et al}
\cite{Nakajima-JPSJ-1993} from a fit to the spin-wave dispersion
of La$_2$NiO$_4$ were $15.5$\,meV and $0.52$\,meV respectively. We
find, therefore, that the spin anisotropy is strongly reduced in
the doped compound, but the strength of the nearest-neighbour
exchange interaction in the antiferromagnetic region is
essentially unaffected by hole-doping.

The lines drawn on Figs.\,\ref{fig:2}(a) and (b) illustrate scan
simulations, and are seen to match the main features of the data
quite well. The inclusion of more exchange parameters might
achieve even better agreement with experiment, but is not
justified given the extent of the present data set. We did,
however, consider an alternative model in which $J'$ couples pairs
of spins displaced by $(a,a)$ across a domain wall instead of
$(2a,0)$. However, the deviations from the first model were slight
and mainly confined to energies close to the zone boundary energy
where the data is sparse. Therefore, while there may be systematic
errors due to the linear approximation and the neglect of more
distant couplings, we expect that the present analysis yields a
good estimate of the relative strengths of the intra- and
inter-stripe spin couplings.

The results presented here provide several new insights into the
dynamics of stripe phases. At high energies ($\gtrsim 30$\,meV)
the spin excitations propagate as if the underlying charge-ordered
lattice were static. A similar observation has been made by
Bourges {\it et al} \cite{Bourges-cond-mat-2002}, who also
commented on a lack of significant anisotropy in the spin-wave
velocity in their study of La$_{0.1.69}$Sr$_{0.31}$NiO$_4$. The
advantage of the La$_{5/3}$Sr$_{1/3}$NiO$_4$ compound studied here
is its simple stripe superstructure that permits us to quantify
the anisotropy in the spin-wave dispersion in terms of a
microscopic model. Thus, while our data are consistent with those
of Bourges {\it et al}, our analysis has determined the ratio
$J/J'$ to be close to 2 and uncovered a dramatic reduction in
single-ion spin anisotropy relative to La$_2$NiO$_4$.

Concerning the anomalous broadening observed in the energy range
10--25\,meV, we earlier discussed several pieces of evidence that
lead us to suggest that this broadening arises from a coupling
between spin excitations and the collective motions of the stripe
domain walls. Further to these, we note that the stripe ordering
temperature ($\sim$$240$\,K) in this compound translates to
$\sim$$20$\,meV. In other words, the thermal energy needed to
destroy long-range order of the stripes is comparable to the
energy where we observe the anomalous broadening in the spin
excitations. Zaanen {\it et al} \cite{Zaanen-PRB-1996} discussed
the dynamics of charged domain wall motion in an antiferromagnetic
background on general theoretical grounds, and argued that domain
wall fluctuations will have their strongest influence on the spin
excitation spectrum at low energies. In one reading this is
consistent with our results, but it is interesting that the
anomalous broadening appears to be restricted to a band of
energies below which the spin excitations recover their sharp
profile. A possible explanation is that there exists a
commensurability gap of $\sim$10\,meV for domain wall motions due
to the pinning of the charges to the lattice. If this were the
case, then one might expect the valley feature to be absent from
the spin excitation spectrum of compounds whose stripe period is
incommensurate with the lattice.

Finally, we draw attention to recent calculations of the imaginary
parts of the charge and spin dynamical susceptibilities of an
ordered stripe system described by the Hubbard model
\cite{Kaneshita-JPSJ-2001, Varlamov-Seibold-PRB-2002}. In addition
to transverse spin waves these calculations also predict
longitudinal modes arising from meandering and compressive
movements of the domain walls. We did not observe sharp
phason-like modes of the type predicted, but their weak intensity
relative to the spin wave scattering and the broadening in the
valley region may have precluded their observation. Similar
calculations for the special case of 1/3 doping, incorporating the
experimentally-determined exchange parameters, would be valuable
for interpreting the features we have measured, and would thus
provide a direct test of this theoretical description of stripe
dynamics.

We thank L.-P. Regnault for help with the IN22 experiments, P.
Isla, D. Gonzalez and R. Burriel for sample characterization, and
J. Zaanen and J. Chalker for stimulating discussions. ATB is
grateful to the Institut Laue-Langevin for support during a
3-month visit in 2002. This work was supported by the Engineering
\& Physical Sciences Research Council of Great Britain.


\end{document}